\documentclass[conference]{IEEEtran}
\IEEEoverridecommandlockouts
\usepackage{cite}
\usepackage{amsmath,amssymb,amsfonts}
\usepackage{algorithmic}
\usepackage{graphicx}
\usepackage{textcomp}
\usepackage{xcolor}
\usepackage{amsmath,graphicx,url,times,booktabs,tabularx,amsfonts,siunitx}
\def\BibTeX{{\rm B\kern-.05em{\sc i\kern-.025em b}\kern-.08em
    T\kern-.1667em\lower.7ex\hbox{E}\kern-.125emX}}

\usepackage{pifont}
\newcommand{\cmark}{\ding{51}}%
\newcommand{\xmark}{\ding{55}}%
    
\begin{document}

\title{Textless Streaming Speech-to-Speech Translation using Semantic Speech Tokens\\
}
\author{Jinzheng Zhao$^{\star}$, Niko Moritz$^{\dagger}$, Egor Lakomkin$^{\dagger}$, Ruiming Xie$^{\dagger}$, Zhiping Xiu$^{\dagger}$, Katerina Zmolikova$^{\dagger}$, \\Zeeshan Ahmed$^{\dagger}$,  Yashesh Gaur$^{\dagger}$, Duc Le$^{\dagger}$, Christian Fuegen$^{\dagger}$\thanks{Work was done when Jinzheng Zhao was an intern at Meta}\\\\
\IEEEauthorblockA{$^{\star}$University of Surrey, UK 
}
\IEEEauthorblockA{$^{\dagger}$Meta AI 
}
}

\maketitle

\begin{abstract}

Cascaded speech-to-speech translation systems often suffer from the error accumulation problem and high latency, which is a result of cascaded modules whose inference delays accumulate.
In this paper, we propose a transducer-based speech translation model that outputs discrete speech tokens in a low-latency streaming fashion. This approach eliminates the need for generating text output first, followed by machine translation (MT) and text-to-speech (TTS) systems. The produced speech tokens can be directly used to generate a speech signal with low latency by utilizing an acoustic language model (LM) to obtain acoustic tokens and an audio codec model to retrieve the waveform.
Experimental results show that the proposed method outperforms other existing approaches and achieves state-of-the-art results for streaming translation in terms of BLEU, average latency, and BLASER 2.0 scores for multiple language pairs using the CVSS-C dataset as a benchmark.

\end{abstract}


\section{Introduction}
\label{sec:intro}

Speech-to-speech translation seeks to transform speech in the source language into speech in the target language.
Conventional speech-to-speech translation systems employ a cascaded approach, comprising automatic speech recognition (ASR), machine translation (MT), and text-to-speech (TTS) processes \cite{nakamura2006atr}. However, this approach suffers from error propagation at each stage, and the cumulative latency of each module makes it challenging to achieve low-delay speech translation results.

To address these issues, we employ discrete audio modeling techniques, which have rapidly advanced and can compress high-dimensional audio into low-dimensional tokens \cite{kumar2024high}. Unlike traditional methods that rely on text as an intermediate representation between source and target speech, our approach directly predicts speech tokens in the target language. This training process does not require translated text as a learning target, making it applicable to unwritten languages.

We utilize streaming ASR methods to predict speech tokens, with the most popular approaches being based on connectionist temporal classification (CTC) \cite{graves2006connectionist} and RNN-Transducer\cite{graves2013speech}.
In this study, we opt for RNN-Transducer as the backbone due to its state-of-the-art performance in streaming speech understanding tasks.

Our contributions can be summarized as follows: (1) We propose a textless streaming speech-to-speech translation model. To our knowledge, this is the first work combining RNN-Transducer and speech tokens for streaming speech-to-speech translation. (2) Experimental results demonstrate that the proposed method achieves state-of-the-art on Spanish $\rightarrow$ English (ES $\rightarrow$ EN) and French $\rightarrow$ English (FR $\rightarrow$ EN) pairs, and achieves competitive performance on German $\rightarrow$ English (DE $\rightarrow$ EN) of CVSS-C \cite{jia2022cvss} dataset in terms of BLEU score, average lagging \cite{ma2020simulmt} and BLASER 2.0 score \cite{barrault2023seamlessm4t}. 

\section{Related Work}
\label{sec:format}
\subsection{Discrete Audio Modelling}
Audio tokenization aims to convert audio waveforms to low-dimensional discrete tokens, which is beneficial in transmission. There are two types of audio tokens. The first is acoustic tokens like SoundStream \cite{zeghidour2021soundstream} and Encodec \cite{defossez2022high}, which employ several residual vector quantization (RVQ) layers to compress the audio where the first layer encode the most important information and the remaining layers encodes the residual information. The second is semantic tokens such as Hubert\cite{hsu2021hubert}, wav2vec-2.0 \cite{baevski2020wav2vec} and w2v-BERT \cite{chung2021w2v}.  There are also some work proposing unified tokens which capture both the semantic and acoustic information. SpeechTokenizer \cite{zhang2024speechtokenizer} employs a similar architecture with RVQ and uses semantic distillation to force the first layer to learn semantic information. Other layers are used to learn acoustic information.

The semantic tokens can be reconstructed to audio waveforms in two stages. In MusicLM,  \cite{agostinelli2023musiclm}, acoustic tokens are generated based on the predicted semantic tokens. Then acoustic tokens are converted to waveform through the SoundStream \cite{zeghidour2021soundstream} decoder. MusicGen \cite{copet2024simple} optimizes the codebook interleaving patterns and predicts the acoustic tokens following the delayed pattern. We use a similar idea for audio reconstruction.
Our AcousticLM maps the acoustic tokens based on the predicted semantic tokens. The AcousticLM has a fixed inference buffer length and emits the acoustic tokens in a chunk-wise streaming way. Finally, acoustic tokens are converted to speech waveform through DAC \cite{kumar2024high} model.

\subsection{Textless Speech-to-Speech Translation}
Textless speech-to-speech translation has thrived in the past few years. These methods employ discrete audio tokens as the learning objectives and use vocoders to synthesize speech.
In \cite{lee2021textless}, discrete speech units extracted from an encoder pretrained with self-supervised learning objective like Hubert \cite{hsu2021hubert} and WavLM \cite{chen2022wavlm} are used as the learning targets. PolyVoice \cite{dong2023polyvoice} employs cross-lingual language models to translate source semantic tokens to target semantic tokens first. Then, the unit-to-speech language model takes the source semantic tokens, target semantic tokens, and source acoustic tokens as input, and output target acoustic tokens. AudioPalm \cite{rubenstein2023audiopalmlargelanguagemodel} extended the large-language model to understand and generate speech with audio tokens from the USM encoder \cite{zhang2023googleusmscalingautomatic}.
In \cite{huang2022transpeech}, bilateral perturbation is introduced to mitigate the acoustic multimodality problem and generate acoustic agnostic learning targets.
In \cite{le2024transvip}, a joint encoder-decoder model first translates the source speech to target text and RVQ codes in the first layer. Then a non-causal LM is used to predict the RVQ codes in the remaining layers.
Textless Translatotron is proposed in \cite{li2023textlessS2S}, which is a non-streaming sequence-to-sequence based speech-to-speech translation model that directly predicts discrete speech representations.
StreamSpeech \cite{zhang2024streamspeech} employs a two-pass paradigm to predict the translated text first and then speech units based on the text. It also uses a multi-task learning paradigm which integrates ASR, speech-to-text translation, and speech-to-speech translation into one model. 

\section{Methods}
\label{sec:pagestyle}
\begin{figure}[tbp]
    \centering
    \includegraphics[width=\columnwidth]{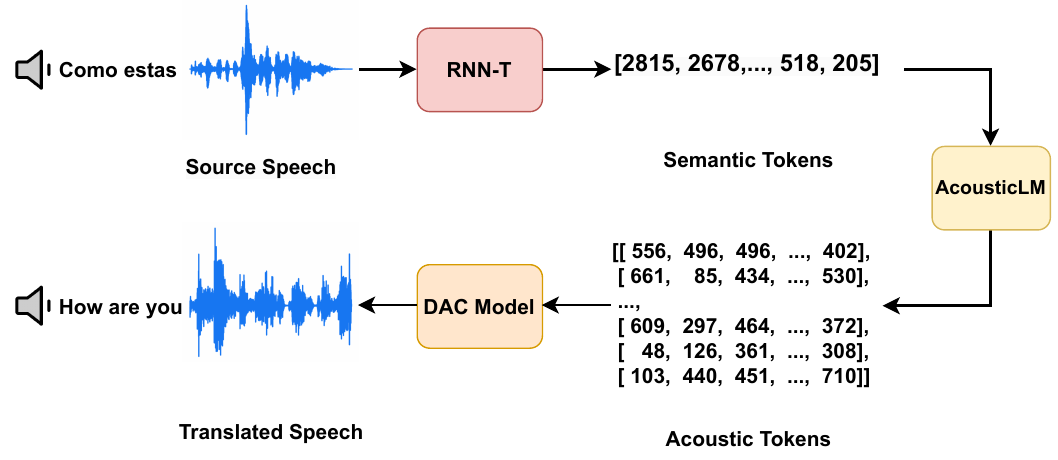}
    \vspace{-10pt}
    \caption{The overall process of generating translated speech with our proposed method. Firstly, the first-pass RNN-T model generates speech tokens at a rate of 25Hz, encoding the spoken content translated into English. Subsequently, the AcousticLM maps these audio tokens into low-level audio tokens, which are then converted into a waveform using a DAC decoder.}
    \label{dac}
    \vspace{-10pt}
\end{figure}

The overall process for our proposed method is demonstrated in Fig. \ref{dac}, including the RNN-Transducer, AcousticLM and DAC model. In addition, we use a pretrained tokenizer to extract the semantic tokens. In this section, we describe each module in detail.
\subsection{RNN-Transducer}

\begin{figure}[tbp]
    \centering
    \includegraphics[width=0.65\columnwidth]{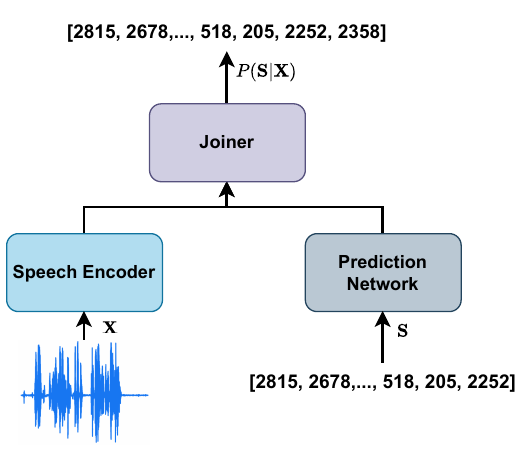}
    \vspace{-10pt}
    \caption{The architecture of the RNN Transducer directly producing speech tokens for speech-to-speech translation.}
    \vspace{-10pt}
    \label{rnnt}
\end{figure}

Let $\mathbf{X} = (\mathbf{x}_{1}, \mathbf{x}_{2}, ..., \mathbf{x}_{T})$ denote the speech features and $\mathbf{S} = (\mathbf{s}_{1}, \mathbf{s}_{2}, ..., \mathbf{s}_{M})$ denote the semantic tokens, where $T$ and $M$ are the number of frames and semantic tokens, respectively.
To achieve streaming textless speech-to-speech translation, we employ the RNN-Transducer model \cite{graves2012sequence}, which takes acoustic features as input and generates semantic tokens extracted from the target speech as output.
The RNN-Transducer loss function determines the alignment between input features and output tokens by marginalizing over a lattice of all possible alignments.
Despite its monotonic alignment properties, previous studies have shown that RNN-Transducer is well-suited for speech-to-text translation tasks \cite{wang2022lamassu, liu2021}, motivating its application in this work.
As depicted in Fig. \ref{rnnt}, the model architecture consists of a speech encoder, prediction network, and joiner.
The speech encoder processes input features into a sequence of embeddings, while the prediction network generates token embeddings based on previously generated tokens.
The joiner predicts the probability of semantic tokens or a blank token based on the output of the encoder and prediction network.
A blank token is generated before the next input feature must be consumed.
The training objective is to minimize the loss function defined as:
\begin{equation}
    L = -\log P(\mathbf{S}|\mathbf{X})
\end{equation}

\noindent where $P(\mathbf{S}|\mathbf{X}) = \sum_{\mathbf{\hat{S}}}P(\mathbf{\hat{S}}|\mathbf{X})$. $\mathbf{\hat{S}}$ denotes all possible alignments between $\mathbf{X}$ and $\mathbf{S}$. 

\subsection{Speech Tokenizer}

The audio semantic tokenizer is designed to capture high-level acoustic information from the speech signal and filter out low-level redundant information. The architecture is inspired by \cite{zhang2023google}, utilizing a quantizer \cite{chiu2022self} loss to extract contextualized discrete representations with a single codebook of size 4,096 tokens.
The tokenizer is trained on English speech data, as our focus in this paper is on translating from other languages into English. 

\subsection{AcousticLM and DAC Model}

The AcousticLM is utilized to convert semantic tokens into acoustic tokens, while the DAC model is employed for reconstructing acoustic tokens into speech waveforms. The acoustic tokens are extracted using a residual vector quantizer (RVQ) \cite{defossez2022high}, which yields a hierarchical structure where the first layer contains the most crucial information and subsequent layers capture remaining details from previous steps.
The AcousticLM comprises stacked Llama layers that take semantic tokens as input and predict acoustic tokens in a delayed pattern \cite{copet2024simple}.

Semantic and acoustic tokens are temporally aligned, which enables use to generate acoustic tokens continuously from small chunks of semantic tokens using the AcousticLM.

\begin{table*}[tbp]
\caption{Experimental results in terms of BLEU and Average Lagging (AL) on CVSS-C dataset. For StreamSpeech, $C$ denotes the chunk size, as defined in \cite{zhang2024streamspeech}. Results for all systems are reported with beam size of 10 in the decoding process.}
\centering
\setlength{\tabcolsep}{3mm}{
\begin{tabular}{lccccccc}
\toprule
& &\multicolumn{2}{c}{\textbf{ES $\rightarrow$ EN}}                        & \multicolumn{2}{c}{\textbf{FR $\rightarrow$ EN}} & \multicolumn{2}{c}{\textbf{DE $\rightarrow$ EN}} \\ \cmidrule(lr){3-4}\cmidrule(lr){5-6} \cmidrule(lr){7-8}
& streaming &\multicolumn{1}{l}{BLEU $\uparrow$} & \multicolumn{1}{l}{AL (ms) $\downarrow$} & BLEU $\uparrow$   & AL (ms) $\downarrow$                & BLEU $\uparrow$         & AL (ms) $\downarrow$        \\ \hline\specialrule{0em}{1pt}{1pt}

S2U \cite{lee2021direct}    &   \xmark        & 18.53                    & -                           & 22.23  & \multicolumn{1}{c}{-}  & 2.99         & -               \\ 
Translatotron 2 \cite{jia2022translatotron} & \xmark    & 22.93                    & -                           & 26.07  & \multicolumn{1}{c}{-}  & 16.91         & -               \\ 

UnitY \cite{inaguma2022unity}     &   \xmark        & 24.95                    & -                           & 27.77  & \multicolumn{1}{c}{-}  & 18.19         & -               \\ 
StreamSpeech ($C$=8) \cite{zhang2024streamspeech} & \cmark & 20.06                    & 1522                        & 22.89  & 1269                   & 14.56         & 1687            \\ 
StreamSpeech ($C$=16) \cite{zhang2024streamspeech} &\cmark& 21.68                    & 2514                        & 24.41  & 2326                   & 15.83         & 2561            \\ 
StreamSpeech ($C$=24) \cite{zhang2024streamspeech} &   \cmark     & 22.36                    & 2999                        & 25.00  & 2803                   & 16.34         & 2978            \\ 
StreamSpeech ($C$=32) \cite{zhang2024streamspeech}  & \cmark      & 22.76                    & 3410                        & 25.20  & 3146                   & 16.57         & 3276            \\ \hline\specialrule{0em}{1pt}{1pt}
our RNN-T (speech-to-text)     & \xmark   & 25.43                    & -                           & 28.40  & - & 22.07         & -               \\ \hline\specialrule{0em}{1pt}{1pt}
our RNN-T (speech-to-speech) (buffer = 10)          & \cmark&24.64                    & 1558                        & 24.72  & 1436                   &   7.65            & 1589                \\ 
our RNN-T (speech-to-speech) (buffer = 30)          & \cmark&25.14                    & 2318                        & 25.27  & 2236                   & 15.98              & 2389                \\ 
our RNN-T (speech-to-speech) (buffer = 50)          & \cmark&25.22                    & 3158                        & 25.36  & 3006                   & 16.20              &  3189               \\ \bottomrule
\end{tabular}}
\label{main}
\vspace{-2mm}
\end{table*}

\begin{table}[tbp]
\caption{Experimental results in terms of BLASER 2.0 REF and QE scores on CVSS-C dataset.}
\centering
\begin{tabular}{lcccccc}
\toprule
& \multicolumn{2}{c}{\textbf{ES$\rightarrow$EN}} & \multicolumn{2}{c}{\textbf{FR$\rightarrow$EN}} & \multicolumn{2}{c}{\textbf{DE$\rightarrow$EN}} \\ \cmidrule(lr){2-3}\cmidrule(lr){4-5} \cmidrule(lr){6-7}
& REF             & QE             & REF & QE             & REF            & QE             \\ \midrule
UnitY (offline)                                                   & 3.26         & 3.33         & 3.20         & 3.18         & 2.94         & 3.15         \\ 
\begin{tabular}[c]{@{}l@{}}StreamSpeech (offline)\end{tabular} & 3.35         & 3.37         & 3.22         & 3.19         & 3.05         & 3.20         \\ \midrule 
\begin{tabular}[c]{@{}l@{}}our RNN-T (buffer = 10)\end{tabular}    & 3.67         & 3.82         & 3.72         & 3.82         & 2.65               &  3.06              \\ 
\begin{tabular}[c]{@{}l@{}}our RNN-T (buffer = 30)\end{tabular}    & \textbf{3.70}         & \textbf{3.84}         & 3.76         & \textbf{3.86}         &  \textbf{3.26}              & \textbf{3.53}               \\  
\begin{tabular}[c]{@{}l@{}}our RNN-T (buffer = 50)\end{tabular}    & \textbf{3.70}         & \textbf{3.84}         & \textbf{3.77}         & \textbf{3.86}         &   \textbf{3.26}             & 3.52               \\ \bottomrule
\end{tabular}
\label{blaser}
\end{table}

\section{Experiments}
\label{sec:exp}

\subsection{Dataset}

We evaluate the model performance on the ES$\rightarrow$EN, FR$\rightarrow$EN, and DE$\rightarrow$EN pairs in CVSS-C \cite{jia2022cvss} benchmark. The source speech is from the CoVoST 2 \cite{wang2020covost} dataset and the target speech is synthesized using a TTS system. 

Due to the limited size of CVSS-C, we first pre-train the system on the Multilingual Librispeech (MLS) dataset \cite{pratap2020mls}. We employ forced alignment with a teacher model to segment the audio into utterances with a maximum duration of 10 seconds.
Since the MLS dataset lacks target speech and only provides sentences in the source language, we utilize an in-house machine translation (MT) system to obtain the text translation into English.
Next, we leverage an in-house TTS system to generate a speech signal of the translated text. Finally, we use the pre-trained speech tokenizer to extract tokens as ground truth. After data preprocessing, we have 466,293 audio clips of ES to EN pairs with an average duration of 7.1 seconds, 547,369 audio clips of FR to EN pairs with an average duration of 7.1 seconds, and 1,002,105 audio clips of DE to EN pairs with an average duration of 7.1 seconds.

\subsection{Evaluation Metrics}
We use ASR-BLEU to evaluate the translation quality, which first transcribes the generated speech using a pretrained ASR system and then calculates the SacreBLEU score \cite{post2018call}. 

For evaluating the latency level of the model, we employ Average Lagging (AL)\cite{ma2020simulmt}, which calculates the difference between the actual emitting time and the ideal emitting time for each token.

We also report the Blaser 2.0 score \cite{barrault2023seamlessm4t} to reflect the translation quality of the generated speech. This metric measures the semantic similarity between the source speech and the target speech through pretrained multilingual encoders \cite{duquenne2023sonar}.

\subsection{Implementation Details}
\label{imple}
\noindent \textbf{Data Preprocessing and Training} The sampling rates for audio clips in MLS and CVSS-C are 16,000 Hz and 48,000 Hz, respectively. We resample the clips in CVSS-C to 16,000 Hz to make the sampling rate consistent. We generate English speech at a sampling rate of 24,000 Hz. We calculate the mel-spectrogram with a window size of 25ms and hop size of 10ms. The feature size is set to 80. Spectrogram augmentation \cite{park2019specaugment} is used with two frequency masks of 27 length and two temporal masks of 100 length with probability of 0.2. We pretrain our model on MLS and fine-tune on CVSS-C with a maximum of 150 epochs with a 5e-4 learning rate and Adam-Sam optimizer. Our model is trained on 32 A100 using fairseq \cite{ott2019fairseq}. 

\noindent \textbf{Model Architecture} For RNN-Transducer, we employ emformer \cite{shi2021emformer} as the encoder. The encoder has a 4-times time reduction layer to match the frame rate of our tokenizer. The emformer has 20 layers with 8 attention heads. It has a segment size of 20 and a right context of 4, which is equal to 800ms and 160ms, respectively. We use LSTM as the prediction layer and a linear layer as the joiner. The output dimension of the joiner is 4097, which matches the token size of our tokenizer plus the blank token. We use beam search for decoding with a beam size of 10. The length penalty is set to 0.5 and 0.05, respectively, for token decoding and text decoding. For AcousticLM, the training buffer size is fixed to 100 tokens. Padded zero will be appended if a smaller inference buffer size is used to lower the latency. The ratio of the acoustic tokens to semantic tokens is 3 and we use 16 layers of acoustic tokens. The frame rate for our tokenizer is 40ms.

\subsection{Baselines}
\noindent\textbf{speech-to-text RNN-T}: We use an RNN-Transducer with the same settings as described in Section~\ref{imple} to train a speech-to-text translation model, i.e. a model outputting a sequence of sentence-piece tokens \cite{kudo2018sentencepiece}. To generate the speech signal, we use an in-house TTS model. We do not report the average lagging results as the used TTS system is non-streaming. 

\noindent\textbf{S2U} \cite{lee2021direct}: An offline system jointly trained by speech2unit prediction and speech2text prediction.

\noindent\textbf{Translatotron 2} \cite{jia2022translatotron}: An offline end-to-end speech-to-text and speech-to-speech system with attention mechanism.

\noindent\textbf{UnitY} \cite{inaguma2022unity}: An offline speech-to-speech translation system that uses a two-pass process to first predict the target text and then the target discrete units. Finally, the predicted discrete units are synthesized to target speech through a vocoder.

\noindent\textbf{StreamSpeech}: Similar to UnitY \cite{inaguma2022unity}, StreamSpeech also has a two-pass process but in a streaming mode. A source CTC decoder for ASR determines the number of input chunks and a target CTC decoder determines the number of output tokens.

\begin{table}[tbp]
\caption{Experimental results of different segment size (seg) and right context (right) on CVSS-C dataset.}
\centering
\setlength{\tabcolsep}{1.5mm}{
\begin{tabular}{cccccccc}
\toprule
&& \multicolumn{2}{c}{\textbf{ES $\rightarrow$ EN}} & \multicolumn{2}{c}{\textbf{FR $\rightarrow$ EN}}& \multicolumn{2}{c}{\textbf{DE $\rightarrow$ EN}} \\ \cmidrule(lr){3-4}\cmidrule(lr){5-6} \cmidrule(lr){7-8}
seg & right & BLEU           & AL(ms)         & BLEU        & AL(ms)               & BLEU                            & AL(ms)               \\ \midrule 
20 & 4 & 25.22          & 3158           & 25.36       & 3006                 & 16.20                           & 3189                 \\ 
32 & 6 & 25.70          & 3419           & 25.79       & 3391                 & 17.78                           & 3410                 \\ \bottomrule
\end{tabular}}
\label{segment}
\vspace{-5pt}
\end{table}

\section{Experiment Analysis}
\label{experiment}
\subsection{Translation Accuracy and Latency}
In Table~\ref{main}, we compare the proposed model to several baselines in terms of translation quality measured by BLEU, and latency measured by Average Lagging. For our model, the latency stems both from the RNN-Transducer and the AcousticLM. For RNN-Transducer, it is determined by the size of the segment and the right context in emformer. For AcousticLM, the latency is mainly affected by the size of a buffer used to convert the semantic tokens to the acoustic tokens. We use different inference buffer sizes in AcousticLM to control the latency. For StreamSpeech baseline, the latency can be controlled by chunk size $C$. 
 
Our speech-to-text translation model achieves the best BLEU score on all language pairs. Compared to StreamSpeech, our model achieves a higher BLEU score on the same level of average latency on ES $\rightarrow$ EN and FR $\rightarrow$ EN pairs. The model performance on ES $\rightarrow$ EN pair outperforms StreamSpeech by an average of 3 BLEU points. The performance is close to the speech-to-text RNN-T and surpasses the offline speech-to-speech translation models such as Translatotron 2 and UnitY.
On FR $\rightarrow$ EN pair, the results of the speech-to-speech RNN-T model are behind the results of speech2text RNN-T model and the offline models. However, it outperforms the streaming baselines, especially on the low-lagging level.
On DE $\rightarrow$ EN pair, our model has competitive performance with the streaming baselines. 

\subsection{Speech Translation Quality}
We present the BLASER 2.0 REF and QE scores in Table \ref{blaser} to assess the semantic similarity between the source speech and target speech. Notably, the QE score is a reference-free metric, whereas the REF score requires the ground truth target speech for evaluation.
The results of StreamSpeech and UnitY are reported as offline baseline methods following \cite{zhang2024streamspeech}. The results of our proposed approach corresponds to streaming speech generation.
Our method achieves a significantly higher BLASER 2.0 score, particularly for ES $\rightarrow$ EN and FR $\rightarrow$ EN translations, indicating superior translation and speech quality.
When examining the model's performance with buffer sizes ranging from 50 to 10, we notice that ASR-BLEU scores decrease, but BLASER 2.0 scores remain relatively stable.
This suggests that the buffer length does not significantly impact the semantic similarity between source and target speech, but it does affect speech quality. Specifically, when calculating the ASR-BLEU score, the ASR system makes more mistakes, resulting in a lower ASR-BLEU score.
An outlier is observed for DE $\rightarrow$ EN when the buffer size is 10, with both BLEU and BLASER 2.0 scores being lower than those of other settings.
The translation for DE $\rightarrow$ EN is more challenging, which can also be observed from the in general lower ASR-BLEU. Therefore, we applied a different version of the speech tokenizer and AcousticLM to better address the relationships between German and English. However, the selected AcousticLM version is more sensitive to small buffer sizes, generating speech of lower quality.

\subsection{Impact of Segment Size and Right Context}
In Table \ref{segment}, we show the impact of using a larger encoder segment and right context size on BLEU score. The results are reported for speech generated with token buffer size of 50 for the AcousticLM. Experimental results show that for ES $\rightarrow$ EN and FR $\rightarrow$ EN, a larger segment size and right context bring a marginal increase in BLEU score with the price of higher latency. For DE $\rightarrow$ EN, there is a large increase in BLEU score. The potential reason is that German has longer words and more reordering compared to the other languages. Increasing the segment size and the right context can make it easier for long-term understanding.  

\section{Conclusion}
\label{con}

We proposed a transducer-based approach to streaming speech-to-speech translation that outputs discrete speech tokens in a low-latency streaming fashion. Our method achieves state-of-the-art results on multiple language pairs of the CVSS-C dataset in terms of BLEU score, average latency, and BLASER 2.0 score. Notably, we demonstrated that our approach can generate high-quality speech while maintaining low latency, making it suitable for real-time applications.
Our findings highlight the promise of transducer-based models directly producing discrete speech tokens for streaming speech-to-speech translation, paving the way for more efficient and effective cross-lingual communication.

\vfill\pagebreak

\begin{bibliography}{ref}
\bibliographystyle{ieeetr}
\end{bibliography}

\end{document}